\documentclass{emulateapj}
\usepackage{apjfonts}

\newcommand{\Msun}      {\mbox{$\rm\,M_{\mathord\odot}$}}

\begin{document}

\lefthead{Truncation of the Accretion Disk}
\righthead{Tomsick et al.}

\submitted{To appear in the Astrophysical Journal Letters}

\def\lsim{\mathrel{\lower .85ex\hbox{\rlap{$\sim$}\raise
.95ex\hbox{$<$} }}}
\def\gsim{\mathrel{\lower .80ex\hbox{\rlap{$\sim$}\raise
.90ex\hbox{$>$} }}}

\title{Truncation of the Inner Accretion Disk around a Black Hole at Low Luminosity}

\author{John A. Tomsick\altaffilmark{1},
Kazutaka Yamaoka\altaffilmark{2},
Stephane Corbel\altaffilmark{3},
Philip Kaaret\altaffilmark{4},
Emrah Kalemci\altaffilmark{5}, and
Simone Migliari\altaffilmark{6}}

\altaffiltext{1}{Space Sciences Laboratory, 7 Gauss Way, 
University of California, Berkeley, CA 94720-7450, USA
(e-mail: jtomsick@ssl.berkeley.edu)}

\altaffiltext{2}{Department of Physics and Mathematics, Aoyama Gakuin
University, Sagamihara, Kanagawa 229-8558, Japan}

\altaffiltext{3}{AIM - Unit\'e Mixte de Recherche CEA - CNRS -
Universit\'e Paris VII - UMR 7158, CEA-Saclay, Service d'Astrophysique,
91191 Gif-sur-Yvette Cedex, France}

\altaffiltext{4}{Department of Physics and Astronomy,
University of Iowa, Van Allen Hall, Iowa City, IA 52242, USA}

\altaffiltext{5}{Sabanci University, Orhanli - Tuzla, Istanbul, 
34956, Turkey}

\altaffiltext{6}{European Space Astronomy Centre, Apartado/P.O. Box 78, 
Villanueva de la Canada, E-28691 Madrid, Spain}

\begin{abstract}

Most black hole binaries show large changes in X-ray luminosity 
caused primarily by variations in mass accretion rate.  An important 
question for understanding black hole accretion and jet production 
is whether the inner edge of the accretion disk recedes at low 
accretion rate.  Measurements of the location of the inner edge 
($R_{\rm in}$) can be made using iron emission lines that arise due 
to fluorescence of iron in the disk, and these indicate that 
$R_{\rm in}$ is very close to the black hole at high and moderate 
luminosities ($\gsim$1\% of the Eddington luminosity, $L_{\rm Edd}$).  
Here, we report on X-ray observations of the black hole GX~339--4 in 
the hard state by {\em Suzaku} and the {\em Rossi X-ray Timing 
Explorer (RXTE)} that extend iron line studies to 0.14\% $L_{\rm Edd}$ 
and show that $R_{\rm in}$ increases by a factor of $>$27 over the 
value found when GX~339--4 was bright.  The exact value of $R_{\rm in}$
depends on the inclination of the inner disk ($i$), and we derive
90\% confidence limits of $R_{\rm in} > 35 R_{g}$ at $i = 0^{\circ}$ and 
$R_{\rm in} > 175 R_{g}$ at $i = 30^{\circ}$.   This provides direct 
evidence that the inner portion of the disk is not present at low 
luminosity, allowing for the possibility that the inner disk is 
replaced by advection- or magnetically-dominated accretion flows.

\end{abstract}

\keywords{accretion, accretion disks --- black hole physics ---
stars: individual (GX~339--4) --- X-rays: stars --- X-rays: general}

\section{Introduction}

Accreting black holes exhibit a variety of spectral states \citep{mr06}, 
most often showing strong thermal emission from an optically thick 
accretion disk \citep{ss73} when they are at their brightest.  As 
luminosities drop below a few percent of the Eddington value 
($L_{\rm Edd}$), the level of thermal emission decreases, and the 
spectrum hardens, signaling a transition to the hard state 
\citep{kalemci04}.  This state is also characterized by having a 
high level of X-ray variability and steady, self-absorbed compact 
jets, which are often observed in the radio band \citep{fender01}.

In theory, it is predicted that the cool gas from the accretion disk 
will evaporate as the mass accretion rate drops \citep{mlm00}.  The 
implied increase in the inner radius of the optically thick disk, 
$R_{\rm in}$, is critical to the idea that a quasi-spherical 
advection-dominated accretion flow (ADAF) forms around the black hole 
\citep{ny94}.  This model requires that $R_{\rm in}$ increases to 
$\sim$100--1000 $R_{g}$ \citep{emn97}, where the gravitational radius 
is $R_{g} = GM/c^{2}$, $G$ and $c$ are constants, and $M$ is the black 
hole mass.

Recently, observational efforts to test the truncated disk picture 
have used two techniques:  Modeling the thermal emission from the
disk and measuring the iron K$\alpha$ emission line in the reflection 
component \citep{lw88}.  The former technique is challenging because 
of uncertainties arising from electron scattering, irradiation of the 
disk, and the torque boundary condition at the disk's inner edge as 
well as the difficulty of modeling a soft X-ray component that is 
strongly affected by interstellar absorption.  Modeling the disk's 
thermal emission to estimate the inner radius ($R_{\rm in}$) has led 
to mixed results with some in favor of an increase in $R_{\rm in}$ 
\citep{gdp08,cabanac09} and others in disagreement 
\citep{rykoff07,rmf09}.  The iron line provides a measurement of 
$R_{\rm in}$ because it is Doppler-broadened due to relativistic motion 
of the accretion disk material and gravitationally redshifted due to 
the black hole's gravitational field.  Broad iron lines have been seen 
from both stellar mass and supermassive black holes 
\citep{tanaka95,fabian09}.  One challenge to the relativistic 
interpretation in the case of stellar mass black holes is that the 
iron line shape may be affected by scattering in a wind 
\citep{lt07,tls09}.  However, arguments in favor of the relativistic 
interpretation are presented in \cite{miller07}.

For stellar mass black holes at high luminosities, the iron line 
profiles have been used extensively, and values of $R_{\rm in}$ 
less than the radius of the innermost stable circular orbit for a 
non-rotating black hole, $R_{\rm in} < 6$~$R_{g}$, have been measured 
for several systems \citep{miller02,miller04,miller08}, including 
GX~339--4.  The broad iron line has, somewhat surprisingly, persisted 
into the hard state \citep{miller06a,tomsick08,hiemstra09}.  However, 
for GX~339--4, the previous detections of the iron line in the hard 
state were obtained at luminosities above 1\% $L_{\rm Edd}$, and it 
remains to be seen if there is evolution in the iron line profile at 
lower luminosities.  

GX~339--4 is one of the most active black hole transients known, 
with 4 major outbursts in the past 7~yr.  The distance to 
GX~339--4 is $\sim$8~kpc \citep{hynes04,zdziarski04}, and the 
orbital period is 1.7~days \citep{hynes03,lc06}.  The lower limit
on the black hole mass is $>$6\Msun~\citep{hynes03,mcm08}, and 
here, we adopt a value of 10\Msun, corresponding to 
$L_{\rm Edd}\sim 1.3\times 10^{39}$ ergs~s$^{-1}$.  GX~339--4 has 
allowed for some of the most in-depth studies of compact jets 
\citep{corbel00,coriat09}.  Here, we use {\em Suzaku} and {\em RXTE} 
observations of GX~339--4 to study the iron line profile at 
0.14\% $L_{\rm Edd}$.

\section{Observations}

We observed GX~339--4 in the X-ray band over a period of 2.2~days 
during 2008 September 24--27 with the {\em Suzaku} \citep{mitsuda07} 
and {\em RXTE} \citep{brs93} satellites.  The observation was made 
1.6~yr after the peak of its 2007 outburst during a time when the 
source was active but faint \citep{russell08,kong08}.  
We analyzed the data using the HEASOFT v6.7 software, and the most 
recent instrument calibrations as of 2009 September 25 for 
{\em Suzaku} and 2009 August 19 for {\em RXTE}.  {\em Suzaku} 
observed GX~339--4 continuously over the 2.2~day period (except 
for Earth occulations), and we accumulated 105,000~s of on-source 
time for the X-ray Imaging Spectrometer (XIS) and 106,800~s for 
the PIN layer of the Hard X-ray Detector (HXD).  The {\em Suzaku} 
data are all contained in observation ID 403067010.  {\em RXTE} 
made five shorter (2,300 to 3,500~s) observations during the same 
time period.  The observation IDs are 93702-04-01-01, 93702-04-01-02, 
93702-04-01-03, 93702-04-02-00, and 93702-04-02-01, and we obtained 
a total Proportional Counter Array (PCA) exposure time of 14,640~s 
and a total High-Energy X-ray Timing Experiment (HEXTE) exposure 
time of 4,625~s.  We also observed GX~339--4 in the radio band
at the Australia Telescope Compact Array (ATCA) on 2008 August 18 
and September 28.  

For observations of relatively faint sources close to the Galactic 
plane, an additional step is required for the PCA spectrum to 
subtract any other emission in its field-of-view \citep{jahoda06}.  
To do this, we obtained data from {\em RXTE} observations of 
GX~339--4 when the source was in quiescence \citep{gfc03}.  These 
observations occurred between 2001 March and 2002 February and 
also on 2003 September 29.  The average quiescent count rate is 
$\sim$6\% of that measured during the 2008 September observations.  
We subtracted the quiescent PCA spectrum for the spectral fits below.

\section{Results}

\subsection{General Properties and Spectral State Identification}

\begin{figure}
\centerline{\includegraphics[width=0.45\textwidth]{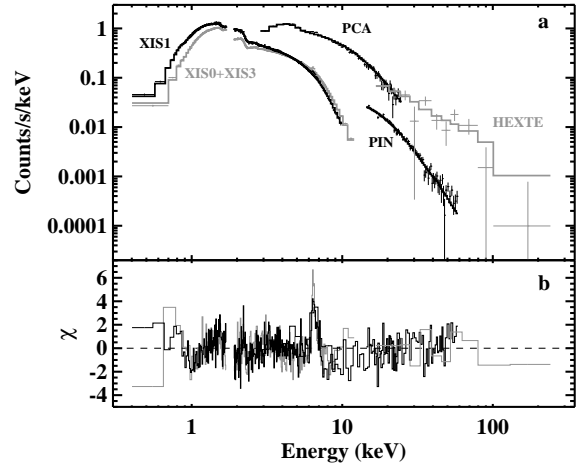}}
%\plotone{fig1.ps}
\vspace{-1.2cm}
\caption{The {\em Suzaku} and {\em RXTE} energy spectrum for 
the entire duration of the 2008 September 24-27 observation of 
GX~339--4.  (a) The counts spectrum fitted with an absorbed 
power-law model.  (b) The residuals for the power-law fit to 
the spectrum given in terms of the contribution from each 
data point to the goodness-of-fit parameter, $\chi$.  The 
largest residuals are seen near 6--7 keV and are due to the 
iron K$\alpha$ emission line.\label{fig:spectrum}}
\end{figure}

We first used the X-ray and radio properties to identify the spectral
state.  We produced the energy spectrum shown in 
Figure~\ref{fig:spectrum}a, and fitted it with an absorbed power-law 
model.  To model the absorption, we used elemental abundances that 
approximate interstellar values \citep{wam00} and atomic cross 
sections from \cite{bm92}.  For the XIS spectra, we did not include 
the data in the 1.7--1.9~keV range due to uncertainties in the 
instrument calibration near the Silicon K-edge.  The overall 
normalizations between instruments were left as free parameters 
in the model, and the fluxes given below are based on those measured 
by XIS0+XIS3.  The measured model parameters are the interstellar 
column density, $N_{\rm H} = (6.79\pm 0.07)\times 10^{21}$ cm$^{-2}$, 
the power-law photon index, $\Gamma = 1.573\pm 0.006$, and the 
unabsorbed 0.4--12~keV flux of $(9.08\pm 0.03)\times 10^{-11}$ 
ergs~cm$^{-2}$~s$^{-1}$ (90\% confidence uncertainties, 
$\Delta\chi^{2} = 2.7$).  Although this model provides a good 
description of the continuum, the quality of the fit is not good 
($\chi^{2}/\nu = 1044/597$) due to the presence of a strong iron 
K$\alpha$ line near 6.4~keV.

To characterize the X-ray timing properties, we produced XIS light
curves in the 0.4--2.5~keV and 2.5--12~keV energy bands with 8~s
time resolution and PCA light curves in the 3--25~keV band with
16~s resolution.  An inspection of the light curves shows that
while they show strong variability and flaring, they are very
similar in the different energy bands, indicating that the spectrum 
changes very little with flux.  We also produced a PCA power 
spectrum at frequencies from 0.005 to 64~Hz.  The power spectrum is 
well-described by a power-law with an index of $1.11\pm 0.03$, and
a fractional rms value of 55\% $\pm$ 2\% (0.01--10~Hz).

The ATCA radio observations show flux densities of $1.10\pm 0.10$ 
and $1.18\pm 0.10$~mJy at frequencies ($\nu$) of 4.80 and 8.64~GHz, 
respectively, on August 18.  For the same frequencies, the 
September 28 flux densities are $0.9\pm 0.2$ and $1.1\pm 0.2$~mJy.
These measurements indicate spectral indices of 
$\alpha = 0.13\pm 0.22$ and $\alpha = 0.3\pm 0.5$, where the 
flux density is given by $S_{\nu}\propto \nu^{\alpha}$.  
Self-absorption of the compact jets that are seen in the hard state 
causes flat or slightly inverted radio spectra ($\alpha\gtrsim 0.0$), 
and our measurements are consistent with this. 

Thus, the hard energy spectrum, the high level of timing noise, 
and the evidence for the presence of a compact jet are all consistent
with the source being in the hard state during our observation.

\subsection{The Iron Emission Line and the Inner Radius of the Disk}

The residuals shown in Figure~\ref{fig:spectrum}b indicate the deviation
of the data from a power-law, and the presence of an iron K$\alpha$ 
emission line is clear.  When we add a Gaussian to model the emission 
line, we measure a line energy of $6.45^{+0.03}_{-0.02}$~keV, consistent 
with neutral to moderately ionized iron, a line width of 
$\sigma = 0.14^{+0.04}_{-0.03}$~keV, and an equivalent width (EW) of 
$77^{+12}_{-10}$~eV.  With the addition of the iron line, the quality
of the fit improves dramatically to $\chi^{2}/\nu = 792/594$.  The 
total 1--100~keV unabsorbed flux is $2.4\times 10^{-10}$ 
ergs~cm$^{-2}$~s$^{-1}$, which is a factor $\sim$9 lower than the 
lowest level at which an iron line was previously detected for 
GX~339--4 \citep{tomsick08}, and this flux corresponds to a luminosity 
of 0.14\% $L_{\rm Edd}$.  

Before discussing the implications of the presence of this narrow iron
line, it is critical to determine if the iron line could be related 
either to poor background subtraction or to emission from other sources 
in the Galactic plane.  For the XIS detectors, we examined the 
background spectrum from two $4^{\prime}.4\times 3^{\prime}.7$ rectangular 
regions on the detectors and verified that only internal background lines 
of the detector appear 
\citep[see][for information about the XIS internal background]{koyama07} 
without any evidence for other background lines with strong emission in 
the iron K$\alpha$ region.

Emission from the Galactic ridge includes iron K$\alpha$ emission with 
the most prominent line being due to He-like iron at 6.7~keV
\citep{koyama86,kaneda97,revnivtsev09}.  Although the emission is very 
strong in the Galactic center region, it decreases for lines-of-sight 
away from the Galactic center and drops especially rapidly with Galactic 
latitude ($b$).  In the Scutum region, at a Galactic longitude of 
$l = 28.5^{\circ}$, a scale height of $0.5^{\circ}$ is estimated
\citep{kaneda97}.  Thus, at the position of GX~339--4 ($l = 338.9^{\circ}$, 
$b =$ --$4.3^{\circ}$), the Galactic ridge emission is expected to be 
weak, but possibly not negligible.

The fact that the XIS0+XIS3 background spectrum does not show an 
emission line at 6.7~keV suggests that we are not detecting Galactic 
ridge emission in our observation of GX~339--4.  As an additional 
check, we produced a new GX~339--4 spectrum using a circular
extraction region centered on the source with a radius of 
$0^{\prime}.86$, which is five times smaller than the $4^{\prime}.3$ 
radius region used previously.  This causes a reduction in the 
source count rate by a factor of 2.1 while reducing the background 
by a factor of 25.  Thus, the strength of any background features 
will decrease by an order of magnitude.  Fitting the new GX~339--4 
XIS spectrum with an absorbed power-law model and inspecting the 
residuals still clearly shows an iron line at 6.4~keV.  Adding a 
Gaussian to fit the iron line, we measure an EW of $73^{+18}_{-14}$~eV, 
which is consistent with no change from the EW of $71^{+11}_{-10}$~eV 
that we measure with XIS with the larger extraction region.  Based 
on this and the background spectrum discussed above, we conclude 
that the iron line is from GX~339--4.

\begin{figure}
\centerline{\includegraphics[width=0.5\textwidth]{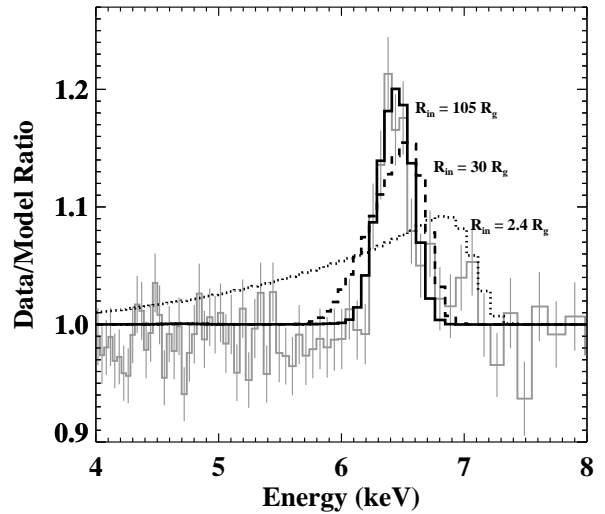}}
%\plotone{fig2.ps}
\vspace{0.0cm}
\caption{Profile of the iron K$\alpha$ emission line from GX~339--4
measured by XIS0+XIS3 (in grey). The profile is shown in terms of a 
data-to-model ratio, where the model is an absorbed power-law 
fitted to the {\em Suzaku} and {\em RXTE} spectra, excluding the 
4.0--8.0~keV region.  We determined the model profiles (in black) 
by restoring the 4.0--8.0~keV data and fitting the entire spectrum 
with a model consisting of an absorbed power-law plus a relativistic 
{\ttfamily laor} iron line.  For the profiles shown, we used a disk 
inclination of $18^{\circ}$ and an emissivity index of $q = 3$.  The 
solid black line shows a profile with $R_{\rm in} = 105$~$R_{g}$, which 
corresponds to the best fit value.  The other two profiles (dashed 
and dotted black lines) were obtained by fixing $R_{\rm in}$ to 
30~$R_{g}$ and 2.4~$R_{g}$, respectively, and then refitting the 
spectrum, allowing the line energy and normalization to adjust.
\label{fig:profile}}
\end{figure}

To use the shape of the iron line to constrain $R_{\rm in}$, we 
return to fitting the full spectrum shown in Figure~\ref{fig:spectrum}, 
and we replaced the Gaussian component with the {\ttfamily laor} 
model \citep{laor91}, which accounts for the relativistic effects 
near a rotating black hole.  The line energy and EW are 
$6.47^{+0.04}_{-0.03}$~keV and $72^{+9}_{-7}$~eV, respectively.  The 
other model parameters are $R_{\rm in}$, the inclination of the inner 
disk, $i$, and the emissivity index, $q$, which is a power-law index 
that sets how the line-emissivity ($J$) of the disk changes with 
radius according to $J\propto r^{-q}$.  Although a broad line allows 
for good constraints on all 3 parameters, this is not the case for a 
narrow line.  Thus, we restricted the range of $q$ to be between 2, 
which corresponds to a value where the contribution to the iron line 
from large radii begins to diverge \citep{laor91}, and 3, which is 
consistent with the value obtained from previous measurements of the 
GX~339--4 iron line in the hard state 
\citep{miller06a,miller08,tomsick08,reis08}.  The inclination has 
been previously measured to be $i = 18\pm 2$ degrees \citep{miller08}.  
For $R_{\rm in}$, the spectral fits indicate 68\% and 90\% confidence 
lower limits of $>$84~$R_{g}$ and $>$65~$R_{g}$, respectively, if 
$i = 18^{\circ}$.  As a value of $R_{\rm in} = 2.4$~$R_{g}$ was obtained 
when the source was bright \citep{miller08}, our results indicate 
that the inner radius changes by a factor of $>$27.  
Figure~\ref{fig:profile} illustrates the huge difference between the 
profile that we measure and a profile with $R_{\rm in} = 2.4$~$R_{g}$.

\begin{figure}
\centerline{\includegraphics[width=0.45\textwidth]{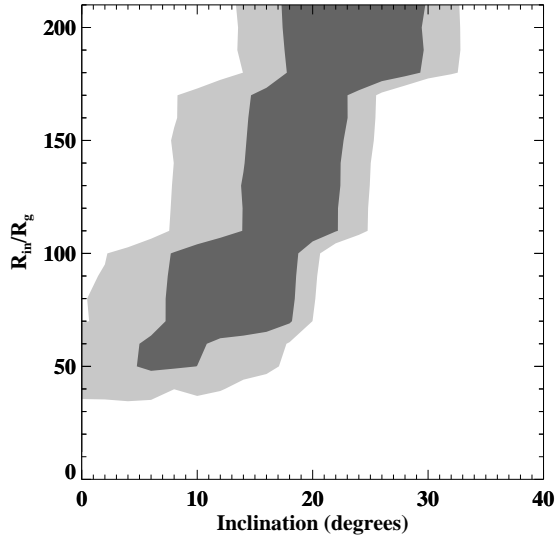}}
%\plotone{fig3.ps}
\vspace{0.0cm}
\caption{Confidence regions for two {\ttfamily laor} iron line
parameters:  The disk inclination and the inner radius of the disk.  
The dark grey region shows the 68\% confidence ($\Delta\chi^{2} = 2.30$)
error region and the light grey region shows the 90\% confidence
($\Delta\chi^{2} = 4.61$) error region.\label{fig:contour}}
\end{figure}

Fixing the disk inclination to $i = 18^{\circ}$ is appropriate for 
constraining the ratio of inner radii at 0.14\% and 12--15\% 
$L_{\rm Edd}$ since we do not expect the disk inclination to change 
significantly with luminosity, being set either by the binary 
inclination or the spin axis of the black hole.  However, for
obtaining a physical value of $R_{\rm in}$, we need to consider the 
inclination.  We refitted the full spectrum with the power-law plus 
{\ttfamily laor} model, allowing $i$ to be a free parameter and $q$ 
to be free in the range 2--3.  Then, we performed a grid search of 
inclinations covering $0^{\circ}$ to $40^{\circ}$ and values of 
$R_{\rm in}$ covering 10 to 210 $R_{g}$.  Figure~\ref{fig:contour} 
shows the results in terms of the 68\% and 90\% confidence contours.  
The contours show that at lower inclinations, the constraint on 
$R_{\rm in}$ becomes somewhat weaker, but even at $i = 0^{\circ}$, 
the results indicate a truncated disk with $R_{\rm in} > 35 R_{g}$ 
(90\% confidence).  On the other hand, at disk inclinations above 
$18^{\circ}$, the inferred values of $R_{\rm in}$ rise rapidly.  For 
example, at $i = 30^{\circ}$, the limit is $R_{\rm in} > 175 R_{g}$.

There are at least two reasons why our limits on $R_{\rm in}$ are
conservative in the sense that the disk may be even more truncated.  
First, \cite{rf05} show that, especially for the disks around 
stellar mass black holes, thermal motions of the disk material can 
lead to significant broadening of the iron line.  If this is the 
case for GX~339--4, then less Doppler broadening from bulk motion
in the accretion disk would be required.  Second, it has been pointed 
out that if the disk inclination is really as low as $18^{\circ}$ and 
the disk inclination is the same as the binary inclination, then 
this would imply a black hole mass of 200\Msun~\citep{cabanac09}, 
which is probably unrealistically high.  Thus, either the disk is 
warped or the inclination is significantly higher than $18^{\circ}$.

\section{Discussion}

\begin{figure}[h]
\centerline{\includegraphics[width=0.45\textwidth]{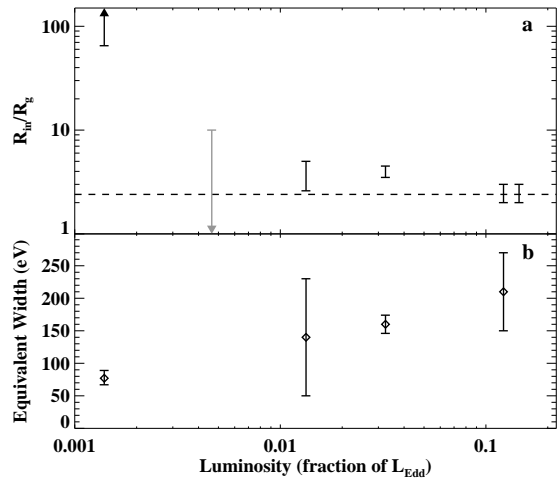}}
%\plotone{fig4.ps}
\vspace{0.0cm}
\caption{Iron line measurements for GX~339--4 over a 0.5--100~keV
luminosity range from 0.14\% to 15\% $L_{\rm Edd}$. 
(a) The constraints on the inner radius of the accretion disk, including 
all GX~339--4 observations made with instruments with energy resolution 
better than 200~eV for which a determination of $R_{\rm in}$ was obtained 
using relativistic reflection models.  The four measurements of 
$R_{\rm in}$ in the range 2--5~$R_{g}$ between 1.3\% $L_{\rm Edd}$ and 
15\% $L_{\rm Edd}$ use {\em Suzaku} \citep{miller08}, {\em XMM-Newton} 
\citep{miller04,miller06a,reis08}, and {\em Swift} \citep{tomsick08}.  
The grey 90\% confidence upper limit is a {\em Swift} measurement 
where broad features in the reflection component are detected, but 
the iron line is not clearly detected \citep{tomsick08}.  The dashed 
line marks the central value obtained at 12--15\% $L_{\rm Edd}$.  The 
measurements indicate that a large increase in $R_{\rm in}$ occurs 
between 0.5\% $L_{\rm Edd}$ and 0.14\% $L_{\rm Edd}$.  (b) The equivalent 
width of the iron line vs. luminosity.  The value we obtain at 0.14\% 
$L_{\rm Edd}$ is the lowest value, which is consistent with a drop in the 
strength of the reflection component due to truncation of the disk. 
\label{fig:r_vs_l}}
\end{figure}

These results provide the most direct and quantitative evidence to 
date for the truncation of the accretion disk for stellar mass black 
holes in the hard state at low luminosities.  Figure~\ref{fig:r_vs_l}a 
compares our constraint on $R_{\rm in}$ at 0.14\% $L_{\rm Edd}$ to previous 
measurements at higher luminosities.  The data show that the inner edge 
of the disk moves sharply outward as the luminosity decreases from 
1\% to 0.1\% $L_{\rm Edd}$.  In addition, the drop in the iron line
EW shown in Figure~\ref{fig:r_vs_l}b is consistent with the increase
in $R_{\rm in}$ since one expects the line to become weaker if the
disk is truncated.  The EW evolution and the fact that the line is 
well-described by a single component provide support for the
interpretation that the narrow line comes from the disk.  Considering 
the broad lines seen in the brighter part of the hard state and the 
disk truncation that we see here, the overall evolution is very 
similar to the theoretical prediction that the disk evaporation will 
start in the middle of the disk and that the entire inner disk will 
evaporate when the luminosity reaches $\sim$0.1\% $L_{\rm Edd}$ 
\citep{taam08}.

While the study of stellar mass black holes, such as GX~339--4,
provides a detailed look at the evolution of $R_{\rm in}$, there 
is iron line evidence for truncated disks around supermassive
black holes in Active Galactic Nuclei (AGN).  For NGC~4258, 
\cite{reynolds09} measure a narrow line with an EW of 45~eV at 
a luminosity of $10^{-3}$\% $L_{\rm Edd}$ and infer that 
$R_{\rm in} > 3\times 10^{3} R_{g}$.  There is also evidence for
a truncated disk in NGC~4593, and for this AGN, there is
possible evidence for a change in $R_{\rm in}$ over a period of 
5~yr \citep{mr09}.

Our finding for GX~339--4 is also important because a large increase 
in $R_{\rm in}$ is required for the presence of an ADAF, and our 
result makes the ADAF model viable for the fainter part of the hard 
state.  However, it is known that the ADAF model does not give a 
complete physical description of the system since it does not 
incorporate the compact jet.  These jets have now been detected 
from GX~339--4 in the hard state both when the disk is truncated 
(this work) and when the disk is not truncated 
\citep{miller06a,tomsick08}.  The possibility that disk truncation 
occurs at a different mass accretion rate than the transition to 
the hard state and the turn-on of the compact jet has not been 
considered in most theoretical work.  This finding strongly 
constrains models such as the magnetically-dominated accretion 
flow (MDAF) model \citep{meier05} where jet production and 
properties depend on the accretion geometry.

%\vspace{0.5cm}

\acknowledgments

We thank Tasso Tzioumis for carrying out the radio observations and 
David M.~Smith for information about {\em RXTE} observations of 
GX~339--4.  We appreciate comments from Steven Boggs and Felix 
Mirabel.  JAT acknowledges partial support from NASA {\em Suzaku} 
Guest Observer grant NNX09AG46G.  EK acknowledges TUB\.ITAK grant 
106T570 and the Turkish Academy of Sciences.  EK and SC acknowledge 
the EU FP7 ITN ``Black Hole Universe.''  This research has made use 
of data obtained from the {\em Suzaku} satellite, a collaborative 
mission between the space agencies of Japan (JAXA) and the USA (NASA). 

% BIBLIOGRAPHY
%\bibliographystyle{jwapjbib}
%\bibliography{refs}

\end{document}